\newcommand\apjl{{\it ApJ}}% Astrophysical Journal, Letters 
\newcommand\apjs{{\it ApJS}}% Astrophysical Journal, Supplement 
\newcommand\ao{{\it Appl.~Opt.}}% Applied Optics 
\newcommand\aap{{\it A\&A}}% Astronomy and Astrophysics 
\newcommand{\e}{\epsilon}

\newcommand{\gp}{\gamma^{\prime}}
\newcommand{\dD}{\delta_{\rm D}}

\input{aipcheck}

\documentclass[
    ,final            % use final for the camera ready runs
  ]
  {aipproc}

\layoutstyle{6x9}

\begin{document}

\title{Nonthermal Synchrotron and Synchrotron Self-Compton Emission from 
       GRBs:  Predictions for {\em Swift} and {\em GLAST} }


\author{Justin D. Finke}{
  address={U.S. Naval Research Laboratory, Code 7653, Washington, DC 20375 },
  altaddress={NRL/NRC Research Associate},
  email={justin.finke@nrl.navy.mil}
}

\iftrue
\author{Charles D. Dermer}{
  address={U.S. Naval Research Laboratory, Code 7653, Washington, DC 20375 },
}

\author{Markus B\"ottcher}{
  address={Astrophysical Institute, Department of Physics and Astronomy, 
           Athens, Ohio 45701},
  %,altaddress={<author1 address>} % additional visiting address
}

\begin{abstract}
Results of a leptonic jet model for the prompt emission and early
afterglows of GRBs are presented.  The synchrotron component is
modeled with the canonical Band spectrum and the synchrotron
self-Compton component is calculated from the implied
synchrotron-emitting electron spectrum in a relativistic plasma blob.
In the comoving frame the magnetic field is assumed to be tangled and
the electron and photon distributions are assumed to be isotropic.
The Compton-scattered spectrum is calculated using the full Compton
cross-section in the Thomson through Klein-Nishina using the Jones
formula.  Pair production photoabsorption, both from ambient radiation
in the jet and from the extragalactic background light (EBL), is taken
into account.  Results are presented as a function of a small set of
parameters: the Doppler factor, the observed variability timescale,
the comoving magnetic field, the peak synchrotron flux, and the
redshift of the burst.  Model predictions will be tested by
multiwavelength observations, including the {\em Swift} and {\em
GLAST} satellites, which will provide unprecedented coverage of GRBs.
\end{abstract}

\maketitle

%%%%%%%%%%%%%%%%%%%%%%%%%%%%%%%%%%%%%%%%%%%%
%% MAINMATTER
%%%%%%%%%%%%%%%%%%%%%%%%%%%%%%%%%%%%%%%%%%%%

\section{Introduction}

The nature of the prompt emission of gamma-ray bursts (GRBs) is still
a mystery.  The most successful theory to date in explaining GRBs and
their afterglows is the fireball model.  In this model, some
mechanism, such as core-collapse supernovae in the case of long GRBs, 
and possibly compact object mergers in the case of short GRBs produces
an expanding shell in a baryon-free environment.  When the expanding
shell collides with another shell or interstellar material, it
produces shocks which accelerate electrons and causes the prompt
emission.  High energy nonthermal electrons can generate such luminous
radiation through synchrotron emission and Compton scattering of
synchrotron photons (synchrotron self-Compton or SSC).  See
\citet{meszaros06} for a recent review.

\begin{figure}
\label{grbfig1}
  \includegraphics[height=.28\textheight]{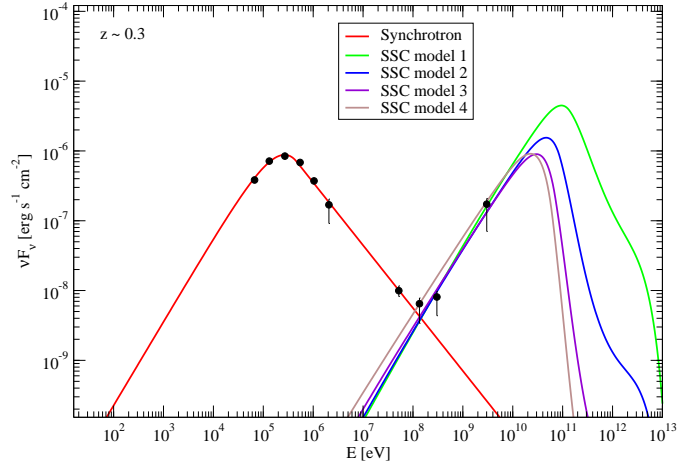}
  \caption{Synchrotron/SSC models for the GRB 940217 assuming $z\sim 0.3$.  }
\end{figure}

\section{One Zone Synchrotron/SSC Model and Fitting Technique}

In the one zone synchrotron/SSC model for GRBs, a portion of a
spherical shell, directed along our line of sight with Doppler factor
$\dD$, is assumed to be filled with nonthermal, isotropically-distributed electrons
and ions, and to entrain a homogeneous, randomly-oriented magnetic field, $B$.
When simulating synchrotron/SSC emission from GRBs, the usual method
is to inject an electron distribution and allow it to evolve, and then
simultaneously fit the entire SED by synchrotron and SSC components
\citep{chiang99}.  This leads to a large number of parameters for the 
modeler to fit.  We have developed a new approach, originally
developed for TeV X-ray selected BL Lacertae objects \citep{finke08}.
In this technique, the low-energy component is assumed to be
synchrotron emission; by fitting this with an empirical function, one
can directly obtain the electron distribution using the
$\delta$-approximation for $\nu F_{\nu}$ synchrotron flux,
\begin{equation}
\label{fesy}
f_\e^{syn} \cong {\dD^4\over 6\pi d_L^2} \; c\sigma_{\rm T} U_B^\prime 
\gamma^{\prime 3}_s N_e^\prime (\gp_s)\ ,
\end{equation}
\citep[e.g.,][]{dermer02} where $d_L$ is the luminosity distance, $c$ 
is the speed of light, $U_B^\prime=B^2/(8\pi)$ is the magnetic field energy
density, $\e$ is the observed dimensionless photon energy,
$\gp_s=\sqrt{ 4\times 10^{13}\e(1+z)/(\dD B)}$ is the electron Lorentz
factor, and $N_e^\prime (\gp_s)$ is the electron distribution.  In the
case of GRBs, the synchrotron spectrum is fit with the empirical
representation of
\citet{band93} which, in $\nu F_{\nu}$ flux, is
\begin{equation}
\label{band}
f_\e^{syn} = \left\{ \begin{array}{ll}
                  f_{peak}\ \left[ \frac{\e}{\e_{peak}} \right]^a\ 
		  \exp\left[-a\left(\frac{\e}{\e_{peak}}-1\right)\right]
		  & \left( \e \le \frac{\e_{peak}(a-b)}{a} \right) \\
		  f_{peak}\ \left[\frac{(a-b)\e_{peak}}{a}\right]^{a-b}\ 
		  \frac{\e^b}{\e^a}\ e^b
		  & \left( \e > \frac{\e_{peak}(a-b)}{a} \right)
		  \end{array}
          \right.
\end{equation}
where $f_{peak} = f_\e^{syn}(\e_{peak})$ and $a$ and $b$ are power-law
indices.  Once $N^\prime_e(\gp)$ is determined, the SSC flux
can be calculated.  We use the full Compton cross-section
in the Thomson through Klein-Nishina regimes based on the formulae of
\citet{jones68}.  Assuming the redshift, $z$, of the burst is known, 
one can model the SSC emission as a function of a small set
of parameters: $\dD$, $B$, and $\Delta R^\prime$, the (comoving) size
scale of the emitting region.  The value of $\Delta R^\prime$ can be
estimated from the observed variability time, $t_{var}$, based on
light travel time arguments by $\Delta R^\prime = \dD c t_{var}/(1 +
z)$, which effectively reduces the number of parameters in the fit to
two.  See \citet{finke08} for more details on this technique.

The ``jet'' power, the power available in the expanding shell that can
create observed radiation, is a combination of the power in the
nonthermal particles and in the magnetic field of the emitting region
\citep{celotti93}.  It gives the upper limit on the observed isotropic
luminosity, and is calculated by
$
L_j = 2 \pi \Delta R^\prime\beta\Gamma^2c\left( U_B^\prime + U_p^\prime \right)
$, 
where $U_p^\prime$ is the particle energy density.  The jet power can also
allow one to constrain the magnetic field, by finding the magnetic
field strength which minimizes this power \citep{dermer04}.

\section{Example:  Fits to GRB 940217}

As an example of this technique, we fit the GRB 940217 with this
synchrotron/SSC model.  The Compton Gamma-Ray Observatory {\em CGRO}
observed this GRB, one of the longest and most energetic GRBs to be
detected at GeV energies.\citep{hurley94}.  We include absorption by
internal jet radiation \citep{gould67} and by the extragalactic
background light \citep{stecker06a}.  The \citet{band93} fit to the
synchrotron spectrum gives the parameters $a=1.2$, $b=-0.9$,
$\e_{peak}=0.53$, and $f_{peak}=8.8 \times 10^{-7}$ erg s$^{-1}$
cm$^{-2}$.  The redshift to the GRB 940217 is unknown; however, it can
be estimated to be $z\sim 0.3$ based on the correlation of
\citet{atteia03}.  This redshift value is used in models 1\ --\ 4
(Fig.\ \ref{grbfig1}), while the value $z\sim 1.0$ is used in models
5\ --\ 8 (Fig.\ \ref{grbfig2}). The parameters for these models can be
seen in Table \ref{modeltable}.  In all models, the variability
timescale was taken to be 1 sec, consistent with {\em CGRO}
observations \citep{hurley94}.  These fits were done in a
``$\chi$-by-eye'' fashion, although for better data, $\chi^2$
minimization can be done \citep{finke08}.

%%%%%%%%%%%%%%%%%%%%%%%%%%%%%%%%%%%%%%%%%%%%
%% Sample figure:
%%
%% The option [height=...] scales the picture to the given height,
%% without it it would be printed at its nominal size
%%%%%%%%%%%%%%%%%%%%%%%%%%%%%%%%%%%%%%%%%%%%

\begin{figure}
\label{grbfig2}
  \includegraphics[height=.28\textheight]{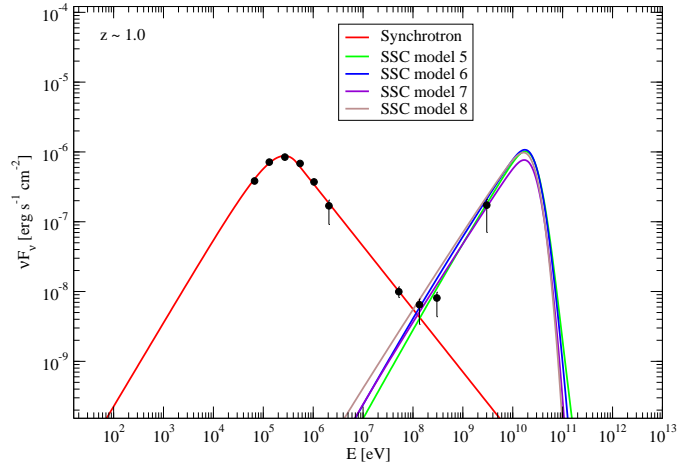}
  \caption{Synchrotron/SSC models for the GRB 940217 assuming $z\sim 1.0$.  }
\end{figure}

%\section{Discussion}

At $z\sim 1.0$, the models are difficult to distinguish from each
other, but at the lower redshift ($z\sim 0.3$), it is possible.  although
all of them give reasonable fits to the {\em CGRO} data.  However, if
the GRB had been observed by {\em GLAST} or a very high energy (VHE)
$\gamma$-ray telescope, it would be possible to distinguish the
models.  Multiwavelength observations of future GRBs by {\em Swift},
{\em GLAST}, and VHE $\gamma$-ray telescopes such as MAGIC, VERITAS,
and HESS will be key in discriminating between models.  TeV
$\gamma$-rays have not been conclusively detected from a GRB so far,
and they would only be expected from nearby GRBs, as farther ones
would have TeV radiation absorbed by the EBL.

At $z\sim 1.0$, the absorption is dominated by the EBL, and internal
photoabsorption effects cannot be seen.  However, the $z\sim 0.3$
models, while still dominated by EBL absorption, show curvature at TeV
energies as a result of internal absorption.  Such curvature in
low-redshift GRBs is one of the key predictions of this model.  

The GRB 940217 has an isotropic luminsosity of $L_{iso} \approx 3.1
\times 10^{50}$ erg s$^{-1}$ if it is located at $z\sim 0.3$, and
$L_{iso} \approx 5.0 \times 10^{51}$ erg s$^{-1}$ if it is at $z\sim
1.0$.  Models which have powers in excess of these can be excluded.

We have summarized a new technique for modeling synchrotron/SSC
broadband spectral energy distributions of GRBs and applied it to the
GRB 940217.  Although the data for this burst is sparse and the
redshift is unknown, this nonetheless illustrates the method.  When
applied to future multiwavelength observations, this technique will be
a good tool for testing the validity of the synchrotron/SSC model.
\boldmath

%%%%%%%%%%%%%%%%%%%%%%%%%%%%%%%%%%%%%%%%%%%%
%% SAMPLE TABLE
%%
%% Shows the use of \tablehead and \tablenote
%% macros
%%%%%%%%%%%%%%%%%%%%%%%%%%%%%%%%%%%%%%%%%%%%

\begin{table}
\begin{tabular}{lrrrr}
\hline
  \tablehead{1}{r}{b}{Model}
  & \tablehead{1}{r}{b}{B [G]}
  & \tablehead{1}{r}{b}{$\delta_D$}
  & \tablehead{1}{r}{b}{$L_j$ [erg s$^{-1}$]}   \\
\hline
\unboldmath
1 & 0.07 & 460 & $1.4 \times 10^{55}$ \\
2 & 1.0  & 370 & $2.8 \times 10^{53}$ \\
3 & 13   & 330 & $6.8 \times 10^{51}$ \\
4 & 141  & 310 & $3.3 \times 10^{50}$ \\
\hline
5 & 0.01 & 930 & $3.0 \times 10^{56}$ \\
6 & 0.87 & 740 & $5.6 \times 10^{54}$ \\
7 & 10   & 680 & $1.4 \times 10^{53}$ \\
8 & 114  & 640 & $6.4 \times 10^{51}$ \\
\hline
\end{tabular}
\caption{Model fit parameters}
%\label{tab:a}
\label{modeltable}
\end{table}

%%%%%%%%%%%%%%%%%%%%%%%%%%%%%%%%%%%%%%%%%%%%%%%%
%% BACKMATTER
%%%%%%%%%%%%%%%%%%%%%%%%%%%%%%%%%%%%%%%%%%%%%%%%

\begin{theacknowledgments}
C.D.D. is supported by the Office of Naval Research.  The work of
J.D.F. is supported by NASA {\em Swift} Guest Investigator Grant
DPR-NNG05ED411 and NASA {\em GLAST} Science Investigation
DPR-S-1563-Y, which also supported a visit by M.B.to NRL.
\end{theacknowledgments}



%\doingARLO[\bibliographystyle{aipproc}]
%          {\ifthenelse{\equal{\AIPcitestyleselect}{num}}
%             {\bibliographystyle{arlonum}}
%             {\bibliographystyle{arlobib}}

\bibliographystyle{aipproc}   % if natbib is available

\bibliography{references,blazar_ref,grb_ref}

%  \typeout{******************************************}
%  \typeout{** Please run "bibtex \jobname" to optain}
%  \typeout{** the bibliography and then re-run LaTeX}
%  \typeout{** twice to fix the references!}
%  \typeout{******************************************}

\end{document}

%%%%%%%%%%%%%%%%%%%%%%%%%%%%%%%%%%%%%%%%%%%
%% The following lines show an example how to produce a bibliography
%% without the help of the BibTeX program. This could be used instead
%% of the above.
%%%%%%%%%%%%%%%%%%%%%%%%%%%%%%%%%%%%%%%%%%%

\endinput
%%
%% End of file `template-6s.tex'.